\documentclass[aps,pra,floatfix,twocolumn,showpacs]{revtex4}
\usepackage{epsfig}
\usepackage{graphicx}
\usepackage{dcolumn}
\begin{document}
\title{Relativistic energy shifts in Ge~II, Sn~II and Pb~II and search 
for cosmological variation of the fine structure constant}
\author{V. A. Dzuba}
\email{V.Dzuba@unsw.edu.au}
\author{V. V. Flambaum}
\email{V.Flambaum@unsw.edu.au}
\affiliation{School of Physics, University of New South Wales, Sydney 2052,
Australia}

\date{\today}

\begin{abstract}
 Sensitivity of
atomic transition frequencies to variation of the fine structure
constant $\alpha=e^2/\hbar c$ increases proportional to $Z^2$
where $Z$ is the nuclear charge. Recently several lines of heavy
 ions Ge~II, Sn~II and Pb~II have been detected in quasar absorption
 spectra. We have performed accurate many-body calculations of
 $\alpha^2$-dependence of transition frequencies
($q$-coefficients) for these atoms 
 and found an order of magnitude increase
in sensitivity in comparison with  atomic transitions
which were previously used to search for temporal and spatial
variation of $\alpha$ in quasar absorption systems.
 An interesting feature in  Pb~II is highly
non-linear dependence on   $\alpha^2$ due to the level pseudo-crossings.

\end{abstract}

\maketitle

\section{introduction}
Theories unifying gravity with other interactions suggest a possibility
of temporal and spatial variation of fundamental  constants of Nature. 
A review of these theories and measurement results can be found
in \cite{uzan}. A very sensitive many-multiplet (MM) method to search for
the variation of the fine structure constant
$\alpha=e^2/\hbar c$  by comparison of quasar absorption spectra with
 laboratory spectra has been suggested in ref \cite{4}.
 It requires calculations
 of relativistic corrections to atomic energy levels
 which allow one to find dependence
of atomic transition frequencies on $\alpha$:
\begin{equation}\label{q}
\omega=\omega_0 + q x
\end{equation}
 where $x=\alpha^2/\alpha_0^2-1 \approx 2 \delta\alpha
/\alpha$.
Here $\omega_0$ and $\alpha_0$ are the laboratory values,
 $\omega$ and $\alpha$ are the rest value of transition
frequency and the fine structure constant for an atom or ion
in a remote cloud of ionized gas located on a distance up to 12 billion light
years from us.  Calculations of $q$ for a number  of atomic
transitions have been performed in Refs. \cite{5,Kozlov,Berengut1,Berengut2}.
 Webb {\textit{et al.}}
 \cite{1,2,3,Murphy}  used MM method \cite{4} and found 
possible evidence of $\alpha$ 
variation, while other 
groups \cite{Srianand,Quast} have used the same method  \cite{4}
but found no indications of the variation ( Note, however,
that works \cite{1,2,3,Murphy} are based on  data from Keck telescope
located in Northern hemisphere while works \cite{Srianand,Quast} used data from
Very Large Telescope located in Southern hemisphere.
Therefore, the disagreement between the results, in principle, may be
attributed to the spatial variation of $\alpha$.).

To improve these results it  would be very important to 
include into data analysis
new atomic  transitions where  possible effects of $\alpha$
variation are larger. Indeed, now systematic effects are at most
comparable to the observed frequency shifts corresponding
to  value of  $\delta\alpha/\alpha=(-0.54 \pm 0.12)\cdot 10^{-5}$
reported in \cite{Murphy}. If transitions with
an order of magnitude larger values of $q$ will be used
these systematic effects will be negligible in comparison
with the frequency shifts produced by 
 $\delta\alpha/\alpha=-0.5\cdot 10^{-5}$.

Coefficients $q$ are small in light atoms and rapidly
increase ($\sim Z^2$) with nuclear charge $Z$. Recently
transitions in heavy ions Ge II ($Z=32$, rest wavelength
1602.4863 A), Sn II ($Z=50$, rest wavelength
1400.450 A)
and Pb II ($Z=82$, rest wavelength
1433.9056 A) have been observed in quasar absorption spectra
\cite{Webb}. Therefore, in the present work we perform calculations
of the dependence of transition frequencies on $\alpha$
in Ge II, Sn II and Pb II. Coefficients $q$ in  Pb II are
an order of magnitude larger than in lighter elements
($Z<31$) which were studied in Refs. \cite{1,2,3,Murphy,Srianand,Quast}. 
Potentially, this may give an order of magnitude increase
of sensitivity to  $\alpha$
variation and help to eliminate
possible disagreement between the results of different
groups.

Note that only E1 transitions from the ground states are
observed in quasar absorption spectra.

\section{Calculations}

Calculations for Ge~II, Sn~II and Pb~II are difficult due to strong
configuration mixing and level pseudo-crossing. As it was pointed out in
our previous works energies of different atomic states considered as
functions of $\alpha^2$ may have very different slopes and come
to crossing at particular values of $\alpha$ \cite{Kozlov}.
Since there can be no real  crossing of levels with the same
conserving quantum numbers the phenomenon is called
level pseudo-crossing. Configurations are strongly mixed and the 
slope of curve representing $E(\alpha^2)$ change very quickly in
the vicinity of level pseudo-crossing. If this happens near
physical value of $\alpha$ ($\alpha \approx \alpha_0$) the results
for the coefficients $q$ (see Eq.~(\ref{q})) become  unstable
since small error in a position of level crossing results in a large
error in $q$.
One way of dealing with this problem is by fitting the experimental values 
of the Land\'{e} $g$-factors \cite{Kozlov}. When states of different total
angular momentum $L$ and total spin $S$ are mixed the resulting values
of $g$-factors depend on configuration mixing similar to
$q$-coefficients ($q \approx \sum q_i W_i$, $g\approx\sum g_i W_i$ where
$W_i$ is the weight of a state $i$). Therefore, correcting the mixing coefficients to
fit the experimental $g$-factors helps to find true values of the
$q$-coefficients as well. However, in the case of ions considered
in present work, there is strong configuration mixing between states
of the same $L,S$ and $J$ ($J$ is the total momentum). 
These states are the $^2D_J$ states of the $ns^2nd$ and $nsnp^2$
configurations ($n=4$ for Ge~II, $n=5$ for Sn~II and $n=6$ for Pb~II).
The mixed states have the same values of $g$-factors and resulting 
$g$-factor does not depend on configuration mixing. In contrast, the
values of $q$-coefficients are still sensitive to configuration mixing.

In principle, it is possible to use experimental hyperfine structure
(hfs) to fit the configuration mixing coefficients. Since single-electron
matrix elements of the hyperfine interaction have very different
values for $s$, $p$ and $d$ states, different configurations must
have different values of the many-electron hfs matrix element.
Therefore, fitting the experimental hfs would have the same effect
as fitting of experimental $g$-factors. Unfortunately, no experimental
data on hfs is available for Ge~II, Sn~II and Pb~II.

This leaves us with energies being the only control of the accuracy
of calculation of the $q$-coefficients (with exception of few levels
of Pb~II where $g$-factors are also useful).
Calculations need to be done
to very high accuracy for the energies to be a reliable control.
The criterion is that deviation of the calculated energies from the
experimental values must be much smaller than the experimental energy
interval between mixed states. The interval between different $^2D$
states of Ge~II, Sn~II and Pb~II is a little larger than 10000~cm$^{-1}$.
Therefore, the deviation of the calculated energies from the
experiment should be less than $\sim$ 1000~cm$^{-1}$ or $<$~1\% of
the excitation energy from the ground state.

A method which can produce the results of desirable accuracy was
suggested in Ref.~\cite{vnm}. Calculations are done in the $V^{N-4}$
approximation. This means that the self-consistent Hartree Fock
procedure is done for the quadruply charged positive ion. 
As it has been demonstrated in Ref.~\cite{vnm}, removal of $s$ and $p$
valence electrons doesn't really affect the atomic core 
(apart from singe-electron energies) and $V^{N-4}$
approximation is a good approximation for all ions with
number of valence electrons ranges from 1 to 5. These include 
neutral atom and negative ion. The main advantage of this
approximation is that core-valence correlations can be relatively
easy included beyond the second-order of the perturbation theory.
As it has been demonstrated in Ref~\cite{vnm} inclusion of
higher-order core-valence correlations can significantly improve
the accuracy of calculations.

The ions of our current interest have three valence electrons and
the effective Hamiltonian for the three-electron wave function of 
valence electrons has the form
\begin{equation}
  \hat H^{\rm eff} = \sum_{i=1}^3 \hat h_{1i} + \sum_{i \neq j}^3 \hat h_{2ij},
\label{heff}
\end{equation}
$\hat h_1(r_i)$ is the one-electron part of the Hamiltonian
\begin{equation}
  \hat h_1 = c \mathbf{\alpha p} + (\beta -1)mc^2 - \frac{Ze^2}{r} + V^{N-4}
 + \hat \Sigma_1.
\label{h1}
\end{equation}
$\hat \Sigma_1$ is the single-electron operator (correlation potential)
which describes correlations between a particular valence electron and 
core electrons. It is the same operator which is used for atoms and ions
with one external electron above closed shells (see, e.g. 
\cite{Dzuba1,Dzuba2,Dzuba3}).
$\hat h_2$ is the two-electron part of the Hamiltonian
\begin{equation}
  \hat h_2 = \frac{e^2}{|\mathbf{r_1 - r_2}|} + \hat \Sigma_2(r_1,r_2),
\label{h2}
\end{equation}
$\hat \Sigma_2$ is the two-electron part of core-valence correlations. 
It represents screening of Coulomb interaction between valence electrons 
by core electrons.

Note that removing $\hat \Sigma$ from the effective Hamiltonian reduces
it to the effective Hamiltonian of the standard configuration
interaction (CI) method. Since we use many-body perturbation theory
to calculate $\hat \Sigma$ the technique we use can be called the
CI+MBPT method \cite{Kozlov2}.

The details of calculations for positive ions of Ge, Sn and Pb will be 
presented elsewhere. Below we discuss specifics of calculations for
Ge~II, Sn~II and Pb~II.

\subsection{Ge~II}

The Ge~II ion is the lightest of three ions ($Z$=32) and the easiest
from computational point of view. The core-valence correlations are relatively 
small due to small number of electrons in the core. The relativistic
 corrections
are small too. The latter means in particular that fine structure is small
and energy multiplets stay well apart from each other and there is no
level pseudo-crossing. Also, there is almost no mixing between states
of different total angular momentum $L$ and/or different total spin $S$.
The $^2D_{3/2,5/2}$ states of the $4s^24d$ and
$4s4p^2$ configurations are still strongly mixed. 
However, due to high accuracy 
of the calculations the final results are very stable.

We calculate $\hat \Sigma_1$ and $\hat \Sigma_2$ for the effective Hamiltonian
(\ref{heff}) in the second order of the MBPT. Inclusion of $\hat \Sigma_1$
brings single-electron energies of Ge~IV to agreement with the experiment
on the level of 0.1\%. No higher-order core-valence correlations need to
be included. The final results are presented in Table~\ref{geii}.
To calculate $q$-coefficients (see Eq.~(\ref{q})) we perform calculations
of the energies for $x = -0.1,0,+0.1$. Than 
$q_+ = 10 (\omega(0.1) - \omega(0)),
 q_- = 10 (\omega(0) - \omega(-0.1))$ and finally $q=(q_++q_-)/2$.
We need both $q_+$ and $q_-$ to check for non-linear behavior of
energies which is usually an indicator of the level pseudo crossing.
For Ge~II $q_+$ and $q_-$ are practically identical for all levels
considered.

\begin{table}
\caption{\label{geii}Energy levels and relativistic energy shift of Ge~II 
(cm$^{-1}$).}
\begin{ruledtabular}
\begin{tabular}{llrrr}
\multicolumn{2}{c}{State} & \multicolumn{2}{c}{Energies} & \multicolumn{1}{c}{$q$} \\
 & & \multicolumn{1}{c}{Expt.\footnotemark[1]} & \multicolumn{2}{c}{Calculations} \\ 
\hline
$4s^24p$ & $^2P^o_{1/2}$ &     0.0 &      0  &     0   \\
         & $^2P^o_{3/2}$ &  1767.1 &   1797  & -1863   \\
			  	     	   		 
$4s4p^2$ & $^4P_{1/2}$   & 51575.5 &  51512  & -3098   \\
         & $^4P_{3/2}$   & 52290.5 &  52241  & -3901   \\
         & $^4P_{5/2}$   & 53366.7 &  53342  & -5001   \\
	
$4s^25s$ & $^2S_{1/2}$   & 62402.4 &  62870  & -664   \\

$4s4p^2$ & $^2D_{3/2}$   & 65015.0 &  65313  & -3387   \\
         & $^2D_{5/2}$   & 65184.1 &  65494  & -3612   \\
			 
$4s^25p$ & $^2P^o_{1/2}$ & 79006.2 &  79386  &  -940   \\
         & $^2P^o_{3/2}$ & 79365.8 &  79750  & -1317   \\

$4s^24d$ & $^2D_{3/2}$   & 80836.1 &  81444  & -2236   \\
         & $^2D_{5/2}$   & 81011.8 &  81625  & -2466   \\
\end{tabular}
\end{ruledtabular}
\noindent \footnotetext[1]{Moore, \cite{moore}}
\end{table}

\subsection{Sn~II}

The Sn~II ion ($Z=50$) is very similar to the Ge~II ion. 
However, correlations and relativistic corrections are larger.
It has some implication on the calculation scheme. It turns out that
inclusion of the higher-order core-valence correlations does lead
to significant improvement of the results for both Sn~IV and Sn~II
ions. We include screening of Coulomb interaction and hole-particle
interaction in all orders of the MBPT. It is done exactly the same way as 
in our calculations for single-valence-electron atoms 
(see, e.g. \cite{Dzuba2}). The $\hat \Sigma_2$ operator is still 
calculated in the second order of the MBPT. 

Final results are presented in Table~\ref{snii}. The $q$-coefficients 
are calculated exactly the same way as for Ge~II. The resulting accuracy 
for energies is very good and the values of $q$-s are very stable.

\begin{table}
\caption{\label{snii}Energy levels and relativistic energy shift of Sn~II 
(cm$^{-1}$).}
\begin{ruledtabular}
\begin{tabular}{llrrr}
\multicolumn{2}{c}{State} & \multicolumn{2}{c}{Energies} & \multicolumn{1}{c}{$q$} \\
 & & \multicolumn{1}{c}{Expt.\footnotemark[1]} & \multicolumn{2}{c}{Calculations} \\ 
\hline
$5s^25p$ & $^2P^o_{1/2}$ &     0.0 &      0  &      0   \\
         & $^2P^o_{3/2}$ &  4251.4 &   4222  &  -4680   \\
			  	     	   		 
$5s5p^2$ & $^4P_{1/2}$   & 46464.2 &  46661  &  -5930   \\
         & $^4P_{3/2}$   & 48368.0 &  48556  &  -8343   \\
         & $^4P_{5/2}$   & 50730.0 &  50915  & -10558   \\
	
$5s^26s$ & $^2S_{1/2}$   & 56885.9 &  56707  &  -2679   \\

$5s5p^2$ & $^2D_{3/2}$   & 58843.8 &  58806  &  -7361   \\
         & $^2D_{5/2}$   & 59463.4 &  59419  &  -8325   \\
			 
$5s^25d$ & $^2D_{3/2}$   & 71405.6 &  71140  &  -5904   \\
         & $^2D_{5/2}$   & 72047.6 &  71804  &  -7015   \\

$5s^26p$ & $^2P^o_{1/2}$ & 71494.3 &  71182  &  -3119   \\
         & $^2P^o_{3/2}$ & 72377.3 &  72061  &  -4071   \\
\end{tabular}
\end{ruledtabular}
\noindent \footnotetext[1]{Moore, \cite{moore}}
\end{table}

\subsection{Pb~II}

The Pb~II ion ($Z=82$) is the most difficult for calculations. 
Correlations are strong and relativistic effects are large too.
Strong $L-S$ interaction lead to intersection of the fine-structure 
multiplets. Also, states of the same total momentum $J$ are
strongly mixed regardless of the values of $L$ and $S$ assigned to them.
The breaking of the $L-S$ scheme can be easily seen e.g. by comparing
experimental values of the Land\'{e} $g$-factors with the non-relativistic
values.

We have done one more step for Pb~II to further improve the accuracy
of calculations as compared to the scheme used for Sn~II. We have
introduced the scaling factors before $\hat \Sigma_1$ to fit the energies
of Pb~IV. These energies are found by solving Hartree-Fock-like
equations for the states of external electron of Pb~IV in the $V^{N-4}$
potential of the atomic core
\begin{equation}\label{Brueck}
  (\hat H_0 + \hat \Sigma_1 - \epsilon_n)\psi_n = 0.
\end{equation}
Here $\hat H_0$ is the Hartree Fock Hamiltonian. $\hat \Sigma_1$ is 
the all-order correlation potential operator similar to what is used for Sn~II.
Inclusion of $\hat \Sigma_1$ takes into account the effect of the core-valence 
correlations on both the energies ($\epsilon_n$) and the wave functions
($\psi_n$) of the valence states producing the so-called Brueckner orbitals.
The difference between Brueckner and experimental energies of the 
$4s$, $4p$ and $4d$ states of Pb~IV are on the level of 0.2 - 0.4\%. 
To further improve the energies we replace $\hat \Sigma_1$
by $f \hat \Sigma_1$ with rescaling factor $f$ chosen to fit the energies
exactly. Then the same rescaled operator $f \hat \Sigma_1$ is used for
the Pb~II ion. It turns out that only small rescaling is needed. Maximum 
deviation of the rescaling factor from unity is  10\%:
$ f(4s)=0.935, \ f(4p_{1/2}) = 1.084, \ f(4p_{3/2}) = 1.1, \
f(4d_{3/2}) = 1.07, \ f(4d_{5/2}) = 1.07. $

The rescaling of the $\hat \Sigma_1$ operator has similar effect on
both ions Pb~IV and Pb~II (and Pb~III) improving agreement with
experiment in all cases. Although, the effect on the energies of Pb~II 
is small, it is important due to the level crossing and strong configuration 
mixing which can make the results to be very unstable if the accuracy
of calculations is not high enough.

Note that we fit the energies of Pb~IV but not Pb~II. The calculations 
for Pb~II can be still considered as pure {\em ab initio} calculations 
since no experimental information about Pb~II is used.

The results are presented in Table~\ref{pbii}. We also present on Fig.~I
the behavior of Pb~II energy levels as functions of $(\alpha/\alpha_0)^2$.
Dotted lines correspond to states with $J=1/2$, short dash lines correspond
to $J=3/2$ and long dash lines correspond to $J=5/2$ (the longer the dash 
the larger the $J$). Horizontal dashes on the right represent experimental
energies.

The resulting picture is very complicated. There are at least three level
pseudo-crossing. One is for the $^2D_{5/2}$ and $^4P_{5/2}$ states at about
$(\alpha/\alpha_0)^2 = 0.8$, another is for the $^4P_{1/2}$ and
$^2S_{1/2}$ states for $(\alpha/\alpha_0)^2 > 1$ and the third one is
for the $^4P_{3/2}$ and $^2D_{3/2}$ states at $(\alpha/\alpha_0)^2 > 1$.
In fact, due to the level pseudo-crossing, these states are so strongly
mixed that their assignment to particular $L-S$ multiplets and configurations
is ambiguous.

To improve the accuracy in $q$ we used experimental information
about the $g$-factors and energy intervals between the strongly mixed levels.
These values are reproduced
with high accuracy in the interval of $(\alpha/\alpha_0)^2$
 between 0.9 and 1. Also, this interval is  far  from the
positions of the  level pseudo-crossings which
 happen at $(\alpha/\alpha_0) > 1$ and 
$(\alpha/\alpha_0)^2 \approx 0.8$. Therefore, the results are
stable. In this situation we
use $q_-$ as the best estimate of the $q$-values:
$q = q_- = 10(\omega(0) - \omega(-0.1))$ ($\omega = \omega(x)$, where 
$x=(\alpha/\alpha_0)^2 -1$). A deviation of ${\overline q}=(q_- + q_+)/2$
from $q_-$ gives us a reasonable accuracy estimate in this case,
$\sim 10\%$. Note again that the energies are calculated with
much higher accuracy. 
\begin{table}
\caption{\label{pbii}Energy levels and relativistic energy shift of Pb~II 
(cm$^{-1}$).}
\begin{ruledtabular}
\begin{tabular}{llrrr}
\multicolumn{2}{c}{State} & \multicolumn{2}{c}{Energies} & \multicolumn{1}{c}{$q$} \\
 & & \multicolumn{1}{c}{Expt.\footnotemark[1]} & \multicolumn{2}{c}{Calculations} \\ 
\hline
$6s^26p$ & $^2P^o_{1/2}$ &     0   &      0  &  0   \\
         & $^2P^o_{3/2}$ & 14081   &  13897  &  -19210   \\
			  	     	   		 
$6s6p^2$ & $^4P_{1/2}$   & 57911   &  58052  &  -21000   \\
         & $^4P_{3/2}$   & 66124   &  66222  &  -34500   \\
         & $^4P_{5/2}$   & 73905   &  73749  &  -40700   \\
	
$6s^27s$ & $^2S_{1/2}$   & 59448   &  59204  &   -6800   \\

$6s^26d$ & $^2D_{5/2}$   & 68964   &  69256  &  -43900   \\
         & $^2D_{3/2}$   & 69740   &  69002  &  -14600   \\

$6s^27p$ & $^2P^o_{1/2}$ & 74459   &  73878  &   -8180   \\
         & $^2P^o_{3/2}$ & 77272   &  76666  &   -12240   \\

$6s6p^2$ & $^2D_{3/2}$   & 83083   &  83196  &  -45200   \\
         & $^2D_{5/2}$   & 88972   &  88800  &  -46700   \\

\end{tabular}
\end{ruledtabular}
\noindent \footnotetext[1]{Moore, \cite{moore}}
\end{table}

\begin{figure}
\centering
\epsfig{figure=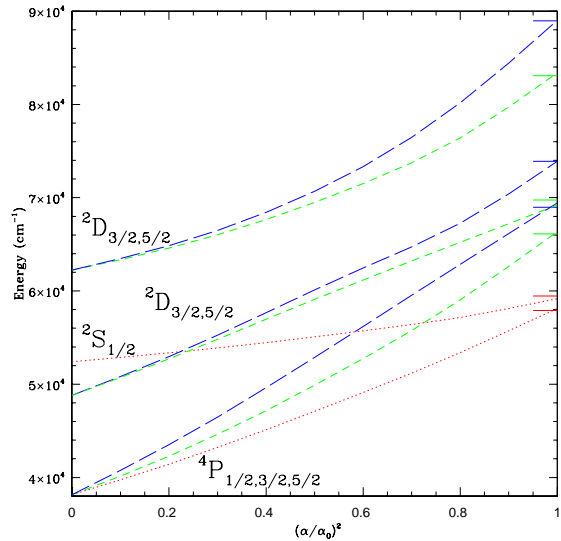,scale=0.38}
\caption{Energies of lowest even-parity multiplets of Pb~II as functions of $\alpha^2$.}
\label{pbalpha}
\end{figure}

\section{Conclusion}
We used combination of the many-body perturbation theory and
 configuration interaction method to calculate dependence of the
energy levels on the fine-structure constant ($q$-coefficients).
This dependence is due to the Dirac relativistic corrections
(which are strongly modified by the many-body effects).
As we found in our previous works the Breit interaction and QED
radiative corrections are not important for the atoms of interest.

   Calculations in Ge II and Sn II are straightforward and 
dependence of the transition frequencies on $\alpha ^2$
 is close to the linear one ($q_+ \approx q_-$).
 The accuracy of the energy level calculation in   Ge II and Sn II is
better than 1\%, for the fine structure intervals the accuracy
is better than 3\%. Therefore, the errors in $q$ should not exceed
3\%. The levels where $q$ are anomalously small may be an exception.
For example, a conservative estimate of accuracy for the level
with $\omega=62402$ cm$^{-1}$ in Ge II where $q=-664$ cm$^{-1}$
is better than 10\%. The value of $q$ for this level ($q=-607$
 cm$^{-1}$) was
also calculated in our previous work \cite{5}. The present calculation
is more accurate.

  The case of Pb II is more complicated due to the level
pseudo-crossing (as functions of $\alpha$). The dependence
on $\alpha^2$ is highly non-linear (see Fig 1). However, 
after taking into account the experimental information about 
the $g$-factors and intervals between energy levels the errors
in $q$-values are reduced to $\sim 10 \%$.

  The values of $q$ in Sn II and especially Pb II are much 
larger than in the elements which were previously used
to search for $\alpha$ variation where $q \sim 1000$ cm$^{-1}$.
For the detected transitions  \cite{Webb}   
 Ge II ($Z=32$, rest wavelength
1602.4863 A) $q=-660$ cm$^{-1}$, Sn II ($Z=50$, rest wavelength
1400.450 A) $q=-5900$ cm$^{-1}$,
and Pb II ($Z=82$, rest wavelength
1433.9056 A) $q=-14600$ cm$^{-1}$. The largest (in absolute value)
 calculated  $q$-coefficient in Pb II is  $q \approx -45000$~cm$^{-1}$.
This may give significant increase in sensitivity
to the variation of $\alpha$. 

\section{Acknowledgments}
We are grateful to J.K. Webb who stimulated this work.
This work is supported by the Australian Research Council.

\end{document}